\documentclass[aps,prd,preprint,groupedaddress]{revtex4-1}
\usepackage{amsmath,graphicx,hyperref,bm}

\begin{document}
\preprint{}

\title{Reference frame dependence of the local measurement of the Hubble constant}

\author{Zhe Chang}
\author{Qing-Hua Zhu}
\email{zhuqh@ihep.ac.cn}

\affiliation{Institute of High Energy Physics, Chinese Academy of Sciences, Beijing 100049, China}
\affiliation{University of Chinese Academy of Sciences, Beijing 100049, China}

\date{\today}

\begin{abstract}
	Observations recently suggest $4.4\sigma$ discrepancy between Hubble constant measured locally and that inferred from cosmic microwave background in standard cosmology.
	Either standard cosmology or local measurement might expect something new.
	We investigate the possibility that the value of Hubble constants might be affected by observers' motional status in the local measurement. Using adapted coordinate for geodesic observers in FLRW space-time and constraints inferred from observation of cosmic shear, we find that the motional status of reference frame could contribute to about $1.1 \pm 0.3\%$ discrepancy of these Hubble constants. Namely,  the difference is $\Delta H_0= 0.81\pm0.22$ km s$^{-1}$ Mpc$^{-1}$, if the locally measured Hubble constant is $ 74$ km s$^{-1}$ Mpc$^{-1}$. This effect seems not negligible as an uncertainty for  local measurement of Hubble constant.
\end{abstract}

\maketitle

\section{Introduction}

Recent local measurement of Hubble constant is discrepant from that inferred
from cosmic microwave background (CMB) up to $4.4\sigma$ \cite{riess_large_2019}. In last decades, the discrepancy has been verified many
times. With $\Lambda$CDM as standard cosmology, CMB \cite{planck_collaboration_planck_2018}, baryon acoustic oscillation \cite{boss_collaboration_cosmological_2015,ross_clustering_2015,alam_clustering_2017,addison_elucidating_2018} and inverse distance ladder technique \cite{aylor_sounds_2019,macaulay_first_2019} provided 
Hubble constants around 67.4 km s$^{-1}$ Mpc$^{-1}$, while local measurements on 
Cepheid variable \cite{riess_redetermination_2009,riess_2.4_2016,riess_milky_2018,efstathiou_h0_2014,zhang_blinded_2017}, gravitational wave \cite{guidorzi_improved_2017,chang_low-redshift_2019,feeney_prospects_2019}, quasar \cite{birrer_h0licow_2019} or tip of the red giant branch \cite{freedman_cosmology_2017,freedman_carnegie-chicago_2019} gave larger Hubble
constants from 69.8 km s$^{-1}$ Mpc$^{-1}$ to 74.03 km s$^{-1}$ Mpc$^{-1}$. The discrepancy of Hubble constants ranges from 1.2$\sigma$ to 4.4$\sigma$.

The local measurement indicates a higher expanding
rate of the current universe. It has inspired cosmological models beyond
$\Lambda$CDM. Most models focused on gravitational repulsion aspect of the universe via a modified dark energy, such as early dark energy \cite{karwal_dark_2016,agrawal_rock_2019,poulin_early_2019} or interaction \cite{pourtsidou_reconciling_2016,di_valentino_can_2017},
phantom-like \cite{di_valentino_reconciling_2016}, phase transition \cite{di_valentino_vacuum_2018,banihashemi_ginzburg-landau_2019} and dynamical aspect of dark energy \cite{zhao_dynamical_2017}. On the other hand, the local measurement of
Hubble constant also can be suffered from systematics error. The error might
be as results of gravitational lensing \cite{smith_effect_2013,grandis_quantifying_2016,kaiser_bias_2016}, vacuum void or density fluctuation \cite{marra_cosmic_2013,ben-dayan_value_2014,odderskov_local_2014,wu_sample_2017,kenworthy_local_2019} and
local gravitational potential \cite{gurzadyan_h_0_2019}. Most of them concluded that it's difficult to alleviate the discrepancy of Hubble constants.

In this paper, we explore the possibility that the discrepancy of Hubble constant  might come from motional status of observers' reference frame.
Namely, observed Hubble constants are different in different reference frames.
We present coordinate transformations between geodesic and static observers in Friedmann-Lemaitre-Robertson-Walker (FLRW) space-time. And it's interesting to find that the Hubble constant in the geodesic frames is  different from that in the isotropic CMB frame.
There is an upper bound of relative deviation of these Hubble constants that can be attributed to motional status of observers' frame, if we assume that redshift space distortions are originated in geodesic motion of heliocentric reference frame.
Namely, using constraints inferred from observation of cosmic shear \cite{qin_bulk_2019}, we find that Hubble constant in the heliocentric reference frame could be larger than that in CMB frame around 1.1\%. In other words, the difference is $0.81\pm0.22$ km s$^{-1}$ Mpc$^{-1}$, if the locally measured Hubble constant is $74$ km s$^{-1}$ Mpc$^{-1}$.
This effect seems considerable compared with the total uncertainty of 1.9\% in the measurement of Hubble constant \cite{riess_large_2019} and might not be neglected.

This paper is organized as follows. In section \ref{II}, we introduce the adapted coordinates of geodesic observers in FLRW space-time. In the geodesic frames, the value of Hubble constant is shown to be different from that in CMB frame. In section \ref{III}, we further clarify the reference frame dependence of Hubble constants via redshift-distance relation in low redshift approximation. In section \ref{IV}, using constraints inferred from observation of cosmic shear, we provide an upper bound of relative deviation of Hubble constants that can be attributed to motional status of heliocentric reference frame. Finally, conclusions and discussions are summarised in section \ref{V}.

\section{Reference frame dependence of Hubble constant}\label{II}

\begin{figure}[ht]
	{\includegraphics[width=0.9\linewidth]{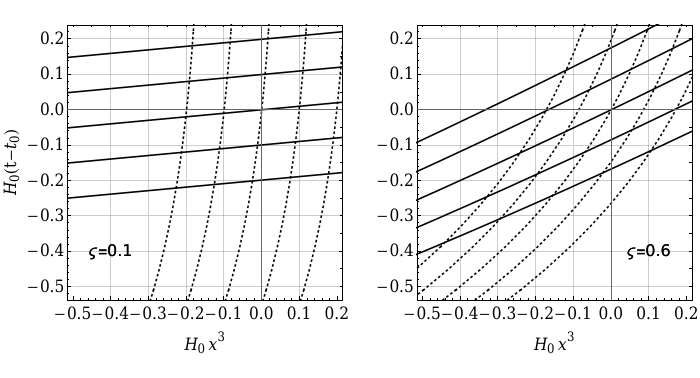}}
	\caption{ $t$--$x^3$ diagram for isotropic FRLW space-time. The $t_0$ is time of the current universe. 
		The dotted and solid lines are coordinate lines of the geodesic reference frames with selected $\varsigma$. Since the diagram is plotted for low-redshift space-time, we set Hubble parameter as a constant for the dark energy dominated epoch. In this case, the scale factor $a$ is in an exponent form. \label{Fig1}}
\end{figure}

In CMB frame, the universe is isotropic and described by the FLRW metric in standard
cosmology,
\begin{equation}
	{\rm{d}} s^2 = - {\rm{d}} t^2 + a^2  (t)(({\rm{d}} x^1)^2 + ({\rm{d}} x^2)^2 + ({\rm{d}}
	x^3)^3)~. \label{1}
\end{equation}
By making up of the metric, the 4-velocities of geodesic observers can be obtained via geodesic equations, $\nabla_u u^{\mu} = 0$,
\begin{equation}
	u^{\mu} = \left( \sqrt{1 + \left( \frac{\varsigma}{a} \right)^2}, 0, 0, \frac{\varsigma}{a^2} \right)~, \label{2-1}
\end{equation}
where $\varsigma$ is an integral constant of geodesic equations and we consider the motions along the direction of axis-$x^3$ for simplicity. 
For given 4-velocities of observers, one can find coordinates $(T,X,Y,Z)$ adapted to the observers, which is formulated as
\begin{equation}
	- N {\rm d}T = u_{\mu} {\rm d}x^{\mu}~, \label{3-1}
\end{equation}
where $N$ is an integrating factor to ensure ${\rm d}^2T=0$. This technique has been involved in 3+1 formalism of general relativity \cite{gourgoulhon_3+1_2012}. In the adapted coordinate $(T,X,Y,Z)$, we require that the geodesic observers should view themselves fixing in spatial coordinates, which is formulated as
\begin{equation}
	\frac{{\rm d} X^i}{{\rm d} \tau} = u^{\mu} \partial_{\mu} X^i = 0~, \label{4-1}
\end{equation}
where $\tau$ is proper time of the 4-velocity $u^{\rm}$ and $X^i$ are spatial coordinates $(X,Y,Z)$ of the geodesic frames. From Eqs.~(\ref{2-1}), (\ref{3-1}) and (\ref{4-1}), we can obtain $N=1$ and
\begin{eqnarray}
	\left\{\begin{array}{lll}
		{\rm d} T = \sqrt{1 + \left( \frac{\varsigma}{a} \right)^2} {\rm d} t - \varsigma {\rm d} x^3~, \\
		{\rm d} X = {\rm d} x^1~,                                                                       \\
		{\rm d} Y = {\rm d} x^2~,                                                                       \\
		{\rm d} Z = \frac{1}{C}\left(-\frac{\varsigma}{a^2 \sqrt{1 + \left( \frac{\varsigma}{a} \right)^2}} {\rm d} t + {\rm d} x^3 \right)~,
	\end{array}\right. \label{5-1}
\end{eqnarray}
where $C$ is an arbitrary constant. From the transformation (Eq.~(\ref{5-1})), one can obtain metric of the adapted coordinates for geodesic observers,
\begin{equation}
	{\rm{d}} s^2 = - {\rm{d}} T^2 + a^2 (T, Z) \left( {\rm{d}} X^2 + {\rm{d}} Y^2 +\left( \frac{1 + \left( \frac{\varsigma}{a} \right)^2}{1 + \varsigma^2}
	\right) {\rm{d}} Z^2 \right) \label{2}~,
\end{equation}
where $a (T, Z) \equiv a (t (T, Z))$. To make sure that it returns to Minkowski space-time in the case of $a \rightarrow 1$, we have set the constant $C = \frac{1}{\sqrt{1+\varsigma^2}}$. 
For the scale factor $a$ beyond 1, one can find that the metric (Eq.~({\ref{2}})) and transformation (Eq.~(\ref{5-1})) are non-trivial. The space-time of the geodesic frames is beyond description of the FLRW metric.  From Eq.~({\ref{5-1}}), we can obtain explicit coordinates via integrals,
\begin{equation}
	\left\{\begin{array}{lll}
		T & = & \int^t_{t_0} \sqrt{1 + \left( \frac{\varsigma}{a (t')} \right)^2}    {\rm{d}} t' - \varsigma  x^3~, \\
		X & = & x^1~,                                                                                           \\
		T & = & x^2~,                                                                                           \\
		Z & = & \sqrt{1 + \varsigma^2} \left( x^3 - \varsigma \int^t_{t_0} \frac{{\rm{d}}
			t'}{a^2 \sqrt{1 + \left( \frac{\varsigma}{a (t')} \right)^2}} \right) ~,
	\end{array}\right. \label{3}
\end{equation}
where $t_0$ is current time in the CMB frame. For the current universe, $a \rightarrow 1$, the transformation turns to be Lorentz transformation. 
The relations between CMB and geodesic frames are shown in Figure \ref{Fig1}. The world lines of geodesic motions are along the dotted lines, which are also coordinate lines of $T$. Here, all the coordinate lines of geodesic frames are shown to be curved in space-time diagram of the CMB frame. This should be understood as effect of the expansion of the universe. 
 
In the geodesic frames (Eq.~(\ref{2})), one might introduce Hubble constants $\bar{H}_0$ by analogy, 
\begin{equation}
	\bar{H}_0 \equiv \left. \frac{\partial_T a}{a} \right|_{a = 1} = \left. \frac{\dot{a} (t)}{a (t)} \frac{\partial t}{\partial T} \right|_{a = 1} = H_0 \sqrt{1 + \varsigma^2} \label{4}~,
\end{equation}
where $H_0 \left( \equiv \left. \frac{\dot{a} (t)}{a (t)} \right|_{a = 1}\right)$ is the Hubble constant in CMB frame.  As we know, it's far from a utility definition of Hubble constant in the geodesic frames for now. However, it's sufficient to suggest that the Hubble constant is reference-frame-dependent and the $\bar{H}_0$ tends to be larger than CMB Hubble constant $H_0$.  

In next section, we would carefully study what's the proper Hubble constant in geodesic frame via distance-redshift relation in low-redshift approximation.

\section{Distance-redshift relation in low-redshifts} \label{III}

In the direction of axis-$Z$, we can obtain equation of trajectories of light rays by making use of ${\rm d}s = 0$,
\begin{equation}
	0 = {\rm d} T \pm \sqrt{\frac{a^2+\varsigma^2}{1+\varsigma^2}} {\rm d} Z~, \label{9-1}
\end{equation}
where $\pm$ represents backward and forward propagating rays, respectively. For a solution of the Eq.~(\ref{9-1}), $F(T,Z)$, we have
\begin{equation}
	{\rm d} F = I(T,Z) \left( {\rm d} T \pm \sqrt{ \frac{a^2 + \varsigma^2}{1 + \varsigma^2} } {\rm d} Z \right)~,
\end{equation}
where
\begin{equation}
	I(T,Z) \equiv \frac{\sqrt{1+\varsigma^2} \mp \varsigma}{\sqrt{a^2+\varsigma^2} \mp \varsigma} ~. \label{11-1}
\end{equation}
The trajectories of light rays propagating from events $(T,Z_c)$ to $(T_0,0)$ satisfy
\begin{equation}
	F(T_0,0) - F(T,Z_c) = 0~. \label{11-2}
\end{equation}
In low-redshift approximation, redshift is measured locally and we can expand Eq.~(\ref{11-2}) for small $Z_c$,
\begin{equation}
	F(T_0,0)-F(T,0)-\left.\frac{\partial F}{\partial Z}\right|_{(T,0)} Z_c + \mathcal{O}(Z_c^2) = 0 ~.
\end{equation}
Thus, we can deduce
\begin{eqnarray}
	Z_c & \approx & \frac{F(T_0,0) - F(T,0)}{\left.\frac{\partial F}{\partial Z_c}\right|_{(T,0)}} \nonumber \\
	& = & \left.\frac{1}{\pm I \sqrt{\frac{a^2+\varsigma^2}{1+\varsigma^2}}} \right|_{Z_c=0}\int^{T_0}_T I(T',0) {\rm d} T'~. \label{13-1}
\end{eqnarray}
From Eqs.~(\ref{5-1}) and the integral along a constant $Z$, we have
\begin{equation}
	{\rm d} Z = \sqrt{1+\varsigma^2} \left(-\frac{\varsigma}{a^2\sqrt{1+\left(\frac{\varsigma}{a}\right)^2}} {\rm d} t + {\rm d} x^3\right) = 0~.
\end{equation}
It leads to
\begin{equation}
	{\rm d} T  =  \frac{{\rm d} t}{\sqrt{1+\left(\frac{\varsigma}{a}\right)^2}} = \frac{{\rm d}a}{a H \sqrt{1+\left(\frac{\varsigma}{a}\right)^2}}~, \label{15-1}
\end{equation}
where $H \equiv \frac{\dot{a}}{a}$. Associating with Eqs.~(\ref{11-1}) and (\ref{15-1}), we rewritten the expression of co-moving distance $Z_c$ (Eq.~(\ref{13-1})) as
\begin{equation}
	Z_c = \pm \left.\frac{1}{\frac{\sqrt{1+\varsigma^2} \mp \varsigma}{\sqrt{a^2+\varsigma^2} \mp \varsigma} \sqrt{\frac{a^2+\varsigma^2}{1+\varsigma^2}}}\right|_{Z_c=0} \int^{\frac{1}{a(T,0)}-1}_0\frac{a {\rm d}\left(\frac{1}{a}-1 \right)}{H \sqrt{1+\left(\frac{\varsigma}{a}\right)^2}} \left\{ \frac{\sqrt{1+\varsigma^2} \mp \varsigma}{\sqrt{a^2+\varsigma^2} \mp \varsigma} \right\}~.
\end{equation}
In low-redshift approximation, $\frac{1}{a(T,0)}-1 \rightarrow 0$, one can obtain
\begin{eqnarray}
	Z_c &=& \pm \frac{1}{H_0 \sqrt{1+\varsigma^2}}\left.\left(\frac{1}{a(T,0)}-1 \right)\right( 1 +    \nonumber \\
	& & + \left.  \frac{1}{2}\left(-1 - q_0 - \frac{\varsigma^2}{1+\varsigma^2} \pm \frac{2 \varsigma}{(1+\varsigma^2)(\sqrt{1+\varsigma^2})} \pm \frac{\varsigma}{\sqrt{1+\varsigma^2}}\right)\left(\frac{1}{a(T,0)}-1 \right) \right. \nonumber \\
	& & \left. + \mathcal{O}\left(\left(\frac{1}{a(T,0)}-1 \right)^2\right) \right)~,
\end{eqnarray}
where the deceleration parameter $q (t) \equiv - \frac{a\ddot{a}}{\dot{a}^2}$.
Likewise, we calculate the co-moving distance in the direction perpendicular to axis-$Z$, for example,
\begin{eqnarray}
	X_c &=& \pm \frac{1}{H_0 \sqrt{1+\varsigma^2}}\left.\left(\frac{1}{a(T,0)}-1 \right)\right( 1 +    \nonumber \\
	& & + \left.  \frac{1}{2}\left(-1 - q_0 - \frac{\varsigma^2}{1+\varsigma^2} \right)\left(\frac{1}{a(T,0)}-1 \right)  + \mathcal{O}\left(\left(\frac{1}{a(T,0)}-1 \right)^2\right) \right)~.
\end{eqnarray}
The difference between co-moving distance $Z_c$ and $X_c$ is originated from Eq.~(\ref{11-1}). For leading order of the co-moving distance $X^{i(0)}_c$, one can find
\begin{equation}
	X_c^{(0)} = Y_c^{(0)} = Z_c^{(0)} = \pm \frac{1}{H_0 \sqrt{1+\varsigma^2}}\left(\frac{1}{a(T,0)}-1 \right)~. \label{19-1}
\end{equation}
In the low-redshift limit, (Eq.~(\ref{19-1})), it seems approaching our claim in section \ref{II} that $\bar{H}_0$ can describe expansion of the universe. However, it's not entire story. In order to obtain distance-redshift relation, we would show how to express $\frac{1}{a(T,0)}-1$ in terms of observed redshifts in the geodesic reference frames.

Using null geodesic equations, we obtain 4-velocities of light rays in the geodesic reference frames,
\begin{eqnarray}
	\frac{{\rm d}T}{{\rm d}\lambda} &=& \frac{\sqrt{(l_1)^2+(l_2)^2+(l_3)^2}\sqrt{a^2+\varsigma^2} - \varsigma l_3}{a^2}~, \\
	\frac{{\rm d}X}{{\rm d}\lambda} &=& \frac{l_1}{a^2}~, \label{21-1}\\
	\frac{{\rm d}Y}{{\rm d}\lambda} &=& \frac{l_2}{a^2}~, \\
	\frac{{\rm d}Z}{{\rm d}\lambda} &=& \frac{\sqrt{1+\varsigma^2}}{a^2}\left(l_3 - \frac{\varsigma \sqrt{(l_1)^2+(l_2)^2+(l_3)^3}}{\sqrt{a^2+\varsigma^2}} \right)~, \label{23-1}
\end{eqnarray}
where $l_1, l_2$ and $l_3$ are integral constants of geodesic equations and can be determined by spatial Killing vectors in the FLRW space-time. For co-moving observers $u^{\mu}=(1,0,0,0)$, the redshift is given by
\begin{eqnarray}
	1 + z & = & \frac{\left. g_{\mu \nu} u^{\mu} \frac{{\rm d}X^{\mu}}{{\rm d}\tau}\right|_{\rm source}}{\left. g_{\mu \nu} u^{\mu} \frac{{\rm d}X^{\mu}}{{\rm d}\tau}\right|_{\rm obs}} \nonumber \\
	& = & \frac{1}{a^2} \left(\frac{\sqrt{a^2+\varsigma^2} - \varsigma \frac{l_3}{\sqrt{(l_1)^2+(l_2)^2+(l_3)^3}}}{\sqrt{1+\varsigma^2} - \varsigma \frac{l_3}{\sqrt{(l_1)^2+(l_2)^2+(l_3)^3}}} \right)~,
\end{eqnarray}
where $g_{\mu \nu}$ is the metric of geodesic reference frames (Eq.~(\ref{2})) and scale factor $a$ is a function of $T$ and $Z$. In low-redshift approximation, we expand the expression of redshift in terms of $\frac{1}{a(T,Z)}-1$,
\begin{equation}
	z = \frac{1+2 \varsigma \left( \varsigma - \sqrt{1+\varsigma^2} \frac{l_3}{\sqrt{(l_1)^2+(l_2)^2+(l_3)^3}} \right)}{1+ \varsigma \left( \varsigma - \sqrt{1+\varsigma^2} \frac{l_3}{\sqrt{(l_1)^2+(l_2)^2+(l_3)^3}} \right)}\left(\frac{1}{a(T,Z)}-1\right) + \mathcal{O}\left(\left(\frac{1}{a(T,Z)}-1 \right)^2\right)~. \label{26-2}
\end{equation}

Firstly, we consider observed redshift of co-moving light sources in the direction of axis-$Z$. In the case of $l_1=l_2=0$, leading order of the redshift in this direction takes the form of
\begin{eqnarray}
	z^3 \equiv z(T,0,0,Z_c) & \approx & \left(1 \pm \frac{\varsigma}{\sqrt{1+\varsigma^2}}\right)\left(\frac{1}{a(T,Z)}-1 \right) \nonumber \\
	& \approx & \left(1 \pm \frac{\varsigma}{\sqrt{1+\varsigma^2}}\right)\left(\frac{1}{a(T,0)}-1 - \left.\frac{\partial_Z a}{a^2}\right|_{Z=0}Z_c   \right) \nonumber \\
	& \approx & \frac{1}{1+ \varsigma^2}  \left(\frac{1}{a(T,0)}-1 \right)~.
\end{eqnarray}
In the third equal, the sign is determined via using Eqs.~(\ref{5-1}) and (\ref{19-1}) and expansion in terms of $a \rightarrow 1$. One can find the redshifts of forward and backward propagating lights are shown to be the same at the leading order.
The second case is observed redshift from co-moving light sources perpendicular to axis-$Z$, (parallel to axis-$X$, for example). It leads to integral constants of light rays, $l_2=0$ and $l_3 = \frac{\varsigma}{\sqrt{1+\varsigma^2}}\sqrt{(l_1)^2+(l_3)^2}$, from which we can obtain the leading order of the redshift in this direction,
\begin{eqnarray}
	z^1 \equiv z(T,X_c,0,0) &\approx& \frac{1+2 \varsigma \left( \varsigma - \sqrt{1+\varsigma^2} \frac{\varsigma}{\sqrt{1+\varsigma^2}}\right)}{1+ \varsigma \left( \varsigma - \sqrt{1+\varsigma^2} \frac{\varsigma}{\sqrt{1+\varsigma^2}} \right)}\left(\frac{1}{a(T,0)}-1\right) \nonumber \\
	&\approx & \frac{1}{a(T,0)}-1~. \label{26-1}
\end{eqnarray}
Eq.~({\ref{26-1}}) would break down in higher redshift, obviously. Once trajectories of the light rays are not approximately straight, we could not require the light rays  both sourced at axis-$X$ and from the direction parallel to axis-$X$. The same situation also exist in calculation of $X_c$. Here, for these light rays, we can estimate its deviation from straight trajectories. Using geodesic equations (Eqs.~(\ref{21-1}) and (\ref{23-1})) and the value of $l_2$, $l_3$, one can obtain
\begin{equation}
	\left|\frac{\Delta Z}{\Delta X}\right| = \frac{\varsigma}{\sqrt{1+\varsigma^2}} \left(\frac{1}{a}-1 \right) + \mathcal{O}\left( \left(\frac{1}{a} -1 \right)^2 \right)~. \label{27-1}
\end{equation}
For redshift survey $z \lesssim 0.03$ \cite{qin_bulk_2019} and $\varsigma \sim 0.3$, it leads to $\left|\frac{\Delta Z}{\Delta X}\right| \lesssim 0.01$. Thus, our calculation above is valid in the low-redshift approximation. 

\begin{figure}[ht]
	{\includegraphics[width=0.9\linewidth]{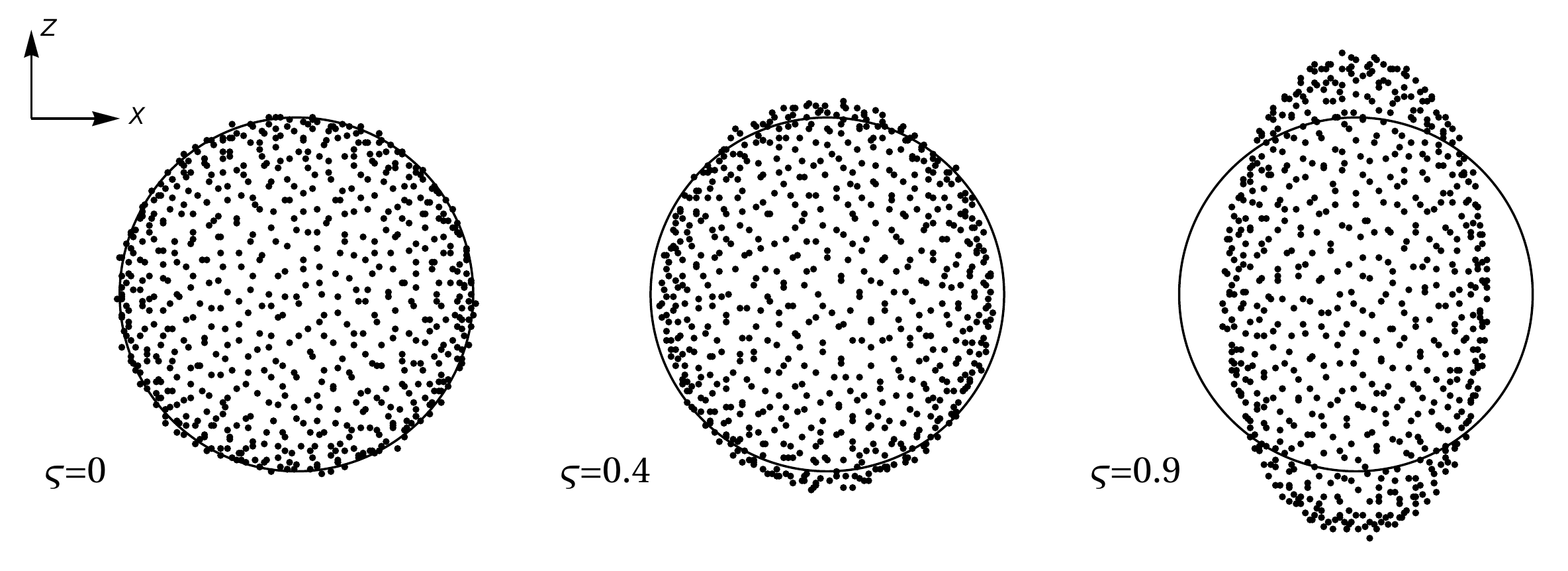}}
	\caption{Schematic diagram for redshift as function of location in low redshift. The parameters $\varsigma$ are selected as $0$, $0.4$ and $0.9$, respectively. For CMB frame in the universe, observed redshifts are the same in a sphere surface of the space. For geodesic frames, the sphere turns to be an ellipsoid. \label{Fig2}}
\end{figure}

From Eqs.~(\ref{19-1}), (\ref{26-1}) and (\ref{27-1}), we can reconstruct distance-redshift relation in the low redshift approximation,
\begin{eqnarray}
	\left\{\begin{array}{lll}
		z^1 = H_0 \sqrt{1+ \varsigma^2} |X_c|~, \\
		z^2 = H_0 \sqrt{1+ \varsigma^2} |Y_c|~, \\
		z^3 = \frac{H_0}{\sqrt{1+\varsigma^2}}   |Z_c|~.
	\end{array}\right. \label{29-1}
\end{eqnarray}
In this approximation, luminosity distance is equal to the co-moving distance $|X_c^i|$. In this distance-redshift relation, there is not difference between forwards and backwards propagating light rays. In Figure \ref{Fig2}, we plot redshift as function of location affected by selected parameters $\varsigma$. Due to geodesic motion of the reference frames, the expansion rates are different in different directions.  

As shown in redshift surveys \cite{strauss_density_1995,watkins_consistently_2009,hong_2mtf_2019,qin_bulk_2019}, the Hubble constant is isotropic part of the expansion in the universe and can be extracted from Eq.~({\ref{29-1}}). It suggests
\begin{eqnarray}
	\left(\begin{array}{c}
			z^1 \\
			z^2 \\
			z^3
		\end{array}\right) &= & H_0 \sqrt{1+\varsigma^2}\left(\begin{array}{ccc}
			1 &   &                           \\
			  & 1 &                           \\
			  &   & \frac{1}{1 + \varsigma^2}
		\end{array}\right) 	 \left(\begin{array}{c}
			|X_c| \\
			|Y_c| \\
			|Z_c|
		\end{array}\right) \nonumber \\
	&= & H_0 \sqrt{1+\varsigma^2} \left( \left( \frac{2}{3} +\frac{1}{3(1+\varsigma^2)} \right)\mathbf{I} + \frac{\varsigma^2}{3(1+\varsigma^2)}\left(\begin{array}{ccc}
				1 &   &    \\
				  & 1 &    \\
				  &   & -2
			\end{array}\right) \right)	 \left(\begin{array}{c}
			|X_c| \\
			|Y_c| \\
			|Z_c|
		\end{array}\right)
	\label{30-1}~,
\end{eqnarray}
and the Hubble constant in geodesic frame should be
\begin{equation}
	\tilde{H}_0 = H_0 \sqrt{1+\varsigma^2}\left( \frac{2}{3} +\frac{1}{3(1+\varsigma^2)} \right)~, \label{31-1}
\end{equation}
where $\mathbf{I}$ is 3-dimensional identity matrix. One can find the expression of the Hubble constant $\tilde{H}_0$ (Eq.~(\ref{31-1})) is different from the CMB $\bar{H}_0$ (Eq.~(\ref{4})). 

In Appendix~\ref{A}, we also present a simpler but less physically-intuitive derivation for Eqs.~(\ref{30-1}).


\section{Estimate parameter $\varsigma$ from cosmic shear and Hubble tension} \label{IV}

The parameter $\varsigma$ is an integral constant of geodesic equations in FLRW space-time. In this section, we would estimate value of the parameter $\varsigma$ from observation. From redshift surveys \cite{strauss_density_1995,watkins_consistently_2009,hong_2mtf_2019}, redshifts of celestial objects are  beyond description of isotropic Hubble flow. 
 As  shown in Ref.~\cite{qin_bulk_2019}, the observed redshift $z_{\rm obs}$ at low-$z$ can be expanded as
\begin{equation}
	z_{\rm obs}(\hat{\bm r}) =H_0 d+ B_i \hat{r_i}+(Q_{i j}\hat{r_i} \hat{r_j})d+\mathcal{O} (d^2)~, \label{8}
\end{equation}
where $z_{\rm obs}$ depends on location of celestial objects $\hat{\bm r}$, $d(=|\bm r|)$ is the distance, vector $B_i$ and traceless tensor $Q_{ij}$ are so-called bulk flow and cosmic shear, respectively. In low redshift approximation, the redshifts are proportional to velocities and can be treated as  vectors with three components in space, namely, so-called redshift space \cite{dodelson_modern_2003},
\begin{equation}
	z_{\rm obs} \equiv z^i_{\rm obs} \hat{r_i} = H_0 d_i \hat{r_i} + B_i \hat{r_i} + Q_{ij} d_j \hat{r_i}~,
\end{equation}
where $d_i \equiv d \hat{r_i}$. One can rewrite the observed redshifts in components,
\begin{equation}
	z^i_{\rm obs} = B_i  + (H_0 \delta_{i j}  + Q_{ij}) d_j~.
\end{equation}
In this point of view, the observed redshifts beyond description of Hubble flow indicate the distortions that appear in redshift space \cite{dodelson_modern_2003}. It has been supported by observation \cite{qin_bulk_2019}. Amplitude of bulk flow is around 300 $ \rm{km\hspace{0.3em} s^{-1}}$ and components of cosmic shear are about 3 $h$ km s$^{-1}$ Mpc$^{-1}$. 

Usually, the redshift space distortions are understood as peculiar velocities $V_{\rm pec} \equiv z_{\rm obs}-H_0 d$, because it's widely accepted that local overdense or void region of the universe can give rise to the peculiar velocity field \cite{dodelson_modern_2003,hong_2mtf_2019,qin_bulk_2019,kenworthy_local_2019}. Conversely, the observations of peculiar velocities could be a probe of the mass distribution in the local universe. On the other side, there is other possibility that the redshift space distortions could arise from motional status of observer's reference frames.  We have shown that it's non-trivial to consider the adapted coordinates for geodesic observers in the universe. Without local structure, the deviations of redshifts beyond Hubble flow might still exist. 

In Eq.~(\ref{30-1}), we have derived redshifts as function of position in geodesic frames, the shear takes the form of
\begin{equation}
{\rm{\bf \tilde{Q}}} = \frac{H_0\varsigma^2}{3\sqrt{1+\varsigma^2}}\left(\begin{array}{ccc}
				1 &   &    \\
				  & 1 &    \\
				  &   & -2
			\end{array}\right) ~,
\end{equation}
and the Hubble constant in the geodesic frame can be expressed as,
\begin{equation}
	\tilde{H}_0 = H_0 \sqrt{1+\varsigma^2}  \left( \frac{2}{3} +\frac{1}{3(1+\varsigma^2)} \right) ~.
\end{equation}
As shown that there is not a dipole in redshift space caused by our geodesic motion, the parameter $\varsigma$ has to be constrained by the cosmic shear. In observation, Qin $et$ $al$ \cite{qin_bulk_2019} presented the cosmic shear ${\rm \bf Q}_{\rm ob}$ measurement for 2MTF, CF3 and the combined data. Here, we neglect peculiar velocities and assume that the observed cosmic shear can at least partly come from the effect of our motion. Namely, from $\sqrt{\tilde{Q}_{i  j} \tilde{Q}^{i  j}} \lesssim (\sqrt{Q_{i	 j} Q^{i  j}})_{\rm obs}$, we can obtain $|\varsigma| \lesssim 0.259 \pm 0.010$. 

Using the value of parameter $\varsigma$, we can estimate how much discrepancy of Hubble constants can be attributed from geodesic motion of heliocentric reference frame. Namely, there is an upper bound of relative deviation of the Hubble constants,

\begin{equation}
	\frac{\tilde{H}_0}{H_0}-1=\frac{2}{3}\sqrt{1+\varsigma^2}+\frac{1}{3\sqrt{1+\varsigma^2}}-1 \lesssim 1.1 \pm 0.3\%~.
\end{equation}
Although,  it seems difficult to alleviate Hubble tension. The effect is considerable compared with the total uncertainty of 1.9\% in recent measurement of Hubble constant \cite{riess_large_2019} and might not be neglected. 

If isotropic part of the redshift is not extracted from redshift space distortions, namely, one might use Eq.~(\ref{29-1}) to test Hubble's law instead of Eq.~(\ref{30-1}), 
the upper bound of the deviation of the Hubble constants could be larger, 
\begin{equation}
	\frac{\bar{H}_0}{H_0}-1=\sqrt{1+\varsigma^2}-1 \lesssim 3.3 \pm 0.2\%~.
\end{equation}
In this case, the Hubble constants can be different in directions.

\section{Conclusions and discussions}\label{V}

In this paper, we explored the possibility that the discrepancy of Hubble constants could be affected by motional status of observers' reference frames. We introduced adapted coordinates for  geodesic observers in FLRW space-time. The geodesic frames are beyond the description of FLRW metric.
If redshift space distortions can be attributed to motion of the geodesic frames, the motional status of observers’ reference frames could contribute to 1.1\% or more riskily 3.3\% 
discrepancy of Hubble constants.  
We can conclude that, firstly, the Hubble constant is, in fact, reference frame dependent. Secondly, as future probes aim at uncertainty of 1\% in measurement of Hubble constants \cite{riess_large_2019}, the effect of observers' motional status should not be neglected. 


In the point of view of peculiar velocities, the bulk flow indicates that most of the celestial objects tend to move towards a specific direction. If the universe is isotropic and homogenous anywhere, it might suggest that our reference frame is moving in the opposite direction of bulk flow with respect to the CMB frame. In this paper, we carefully studied observers' reference frame in the framework of general relativity. The luminosity-redshift relation was calculated in co-moving grids based on the picture that observers are co-moving with the celestial objects in the local group. In this case, there is not difference between forwards and backwards propagating light rays as shown in Eq.~(\ref{23-1}). And the geodesic motion of the reference frames contributes only to the shear part of distortion of redshift space. In this point of view, the cause of bulk flow might be non-geodesic motion, such as accelerated motion of reference frames. It's different from the scenario of peculiar velocities.

Here, the 1.1\% deviation of Hubble constants was inferred from cosmic shear without peculiar velocities. In fact, it would be more realistic to combine the effect of peculiar velocities and the motional status of observers’ reference frames discussed in this paper. Further studies should deal with how to distinguish and entangle these contributions.  

Other key part of our model is estimation of the parameter $\varsigma$ from cosmic shear. 
The upper bound of the parameter $\varsigma$ is determined by assuming that all the observed cosmic shear is from the motion of the geodesic frames. In fact, the $\varsigma$ can be larger, if peculiar velocities might cancel part of the shear. In this case, it would lead to a larger deviation of Hubble constants in difference reference frames. 

Fundamentally, the  work is based on that the adapted coordinates for different geodesic observers are not equivalent. It's interesting on conceptual level. In Minkowski space-time, due to Lorentz symmetry, we can't distinguish between reference frames of static and inertial observers. While in FLRW space-time, it's not true. We can distinguish the reference frames adapted to different geodesic observers. That's the reason that we can figure out the difference of Hubble constant in CMB and heliocentric frame, although there might not be a real observer who is static with respect to CMB frame.

\begin{acknowledgements}
	The authors wish to thank Prof.~Sai Wang, Dr.~Zhi-Chao Zhao, Yong Zhou and Xu-Kun Zhang for discussions. This work has been funded by the National Nature Science Foundation of China  under grant No. 11675182 and 11690022.
\end{acknowledgements}

\appendix
\section{Alternative derivation for Eq.~({\ref{30-1}})}
\label{A}
For co-moving observers $u^{\mu}$, the deviation vector $\xi^{\mu}$ of $u^{\mu}$, which has $[u, \xi]^{\mu} \equiv u^{\nu}\nabla_{\nu} \xi^{\mu} - \xi^{\nu}\nabla_{\nu} u^{\mu}=0$, could indicate shape of space-time. From the property of deviation vectors, one can find
\begin{eqnarray}
	\frac{D}{{\rm d}\tau}\xi^{\mu} = \left(\sigma^{\mu}_{\nu} + w^{\mu}_{\hspace{0.5em}\nu} + \frac{1}{3}\theta\gamma^{\mu}_{\nu}\right)\xi^{\nu}~, \label{A1}
\end{eqnarray}
where $\frac{D}{{\rm d}\tau} \equiv u^{\nu}\nabla_{\nu}$, $\gamma^{\mu}_{\nu} = \delta^{\mu}_{\nu}+ u^{\mu}u_{\nu}$ and
\begin{eqnarray}
	\theta & = & \nabla_{\mu}u^{\mu}~, \\
	\sigma_{\mu \nu} & = &\frac{1}{2}\left(\nabla_{\mu}u_{\nu}+\nabla_{\nu}u_{\mu}  + u_{\mu}\nabla_u u_{\nu} + u_{\nu}\nabla_u u_{\mu}  \right)  - \frac{1}{3}\theta\gamma_{\mu \nu} ~,\\
	w_{\mu \nu} & = & \frac{1}{2} \gamma^{\alpha}_{\mu}\gamma^{\beta}_{\nu}\left( \nabla_{\alpha} u_{\beta} - \nabla_{\beta}u_{\alpha} \right)~.
\end{eqnarray}
The $\theta$, $\sigma_{\mu \nu}$ and $w_{\mu \nu}$ are expansion scalar, shear and rotation tensor, respectively. They are derived from kinematical decomposition of 4-velocity $u^{\mu}$ and so-called kinematical quantities.

In the geodesic reference frame, the 4-velocities of co-moving observers are $u^{\mu} = (1,0,0,0)$. By making use of metric (Eq.~(\ref{2})) of the reference frame, we can obtain
\begin{eqnarray}
	\theta & = & \frac{\partial_T a}{a} \left(2 + \frac{1}{1+ \left(\frac{\varsigma}{a}\right)^2} \right)~, \\
	\sigma_{\mu \nu} & = & \frac{\partial_T a}{a} \frac{\left(\frac{\varsigma}{a}\right)^2}{3\left(1 + \left(\frac{\varsigma}{a}\right)^2 \right)} \left(\begin{array}{cccc}
			0 &   &   &    \\
			  & 1 &   &    \\
			  &   & 1 &    \\
			  &   &   & -2
		\end{array}\right)  ~,\\
	w^{\mu}_{\hspace{0.5em}\nu} & = & 0~.
\end{eqnarray}
In the low-redshift approximation, derivative of deviation vector $\xi^{\mu}$ can be regard as relative velocities between co-moving observers. It suggests $\frac{D}{{\rm d}\tau}\xi^i \approx \upsilon^i = c z^i$. Here, $z^i$ is observed redshift and we set speed of light $c=1$. Thus, the Eq.~(\ref{A1}) in low redshifts can be rewritten as
\begin{equation}
	z^i = \left. \left(\sigma^i_j + w^i_{\hspace{0.5em}j} + \frac{1}{3}\theta\delta^i_j\right) \right|_{a=1}\xi^{j}~.
\end{equation}
This leads to
\begin{eqnarray}
	\left(\begin{array}{c}
			z^X \\
			z^Y \\
			z^Z
		\end{array}\right)
	&= & H_0 \sqrt{1+\varsigma^2}  \left( \left( \frac{2}{3} +\frac{1}{3(1+\varsigma^2)} \right)\mathbf{I} + \frac{\varsigma^2}{3(1+\varsigma^2)}\left(\begin{array}{ccc}
				1 &   &    \\
				  & 1 &    \\
				  &   & -2
			\end{array}\right) \right)	 \left(\begin{array}{c}
			\xi^X \\
			\xi^Y \\
			\xi^Z
		\end{array}\right)~.
\end{eqnarray}
It shows that Eq.~(\ref{30-1}) can be obtained via calculating kinematical quantities of co-moving observers.

\bibliography{citation}
\end{document}